\documentclass[a4paper,fleqn]{cas-dc}

\usepackage[numbers,sort&compress]{natbib} 
\usepackage{graphicx}
\usepackage{dcolumn}
\usepackage{bm}
\usepackage{float}
\usepackage{placeins}
\usepackage{braket}
\usepackage{comment}
\def\tsc#1{\csdef{#1}{\textsc{\lowercase{#1}}\xspace}}
\tsc{WGM}
\tsc{QE}

\begin{document}
\let\WriteBookmarks\relax
\def\floatpagepagefraction{1}
\def\textpagefraction{.001}

\shorttitle{Structure Factor and Dynamics of DMI}

\shortauthors{Wilson, Prescott \& Haraldsen}

\title [mode = title]{General structure factor and dynamic effects of the Dzyaloshinskii-Moriya interaction in S = 1/2 clusters}  

\author[1]{Evan M. Wilson}

\affiliation[1]{organization={University of North Florida},
            addressline={Department of Physics}, 
            city={Jacksonville},
            postcode={32224}, 
            state={FL},
            country={US}}

\author[1]{Joseph A. Prescott}

\author[1]{Jason T. Haraldsen}
\cormark[1]



\begin{abstract}
Understanding the effects of the Dzyaloshinskii-Moriya interaction (DMI) has become increasingly important in the context of nanoscale magnetism and spintronics. In this study, we derive a general structure factor equation for an S = 1/2 dimer and show that the anisotropic ratio $D_z/|J|$ and complex phase $\phi$ of the DMI control the gap energy and intensity of the $|0,0\rangle \to |1,0\rangle$ transition. {Using exact diagonalization of the Heisenberg spin-spin Hamiltonian that incorporates both isotropic and anisotropic interactions,} as well as the effects of an external magnetic field and an electric field. Our results show that the DM interaction splits energy eigenstates, induces level repulsion, and significantly modifies the spin dimer structure factor. These effects reveal a direct correspondence between thermodynamic anomalies in the heat capacity and spin-resolved selection rules.
\end{abstract}


\maketitle

\section{Introduction}\label{Intro}

The study of magnetism has continually unveiled intricate interactions that govern the behavior of spin systems, offering profound insights into quantum systems and material science \cite{Vargova2022, YangAPL24}. Among these interactions, Dzyaloshinskii-Moriya interactions (DMI) have emerged as a crucial mechanism that shapes non-collinear spin structures, spin dynamics, and exotic topological states\cite{dzyaloshinskii1958thermodynamic,MoriyaPR60, bogdanov89JETP, Ham2021, YangNRP2023, Matthies24PRR}. Theoretical and experimental studies have demonstrated that Dzyaloshinskii-Moriya interactions (DMIs) play a fundamental role in stabilizing chiral magnetic textures\cite{Katsura05PRL,Muhlbauer09Sci}, such as Néel skyrmions in multiferroic and transition metal systems\cite{Borisov2024, PanPRB24}. The competition between dipolar interactions and DMIs has been shown to influence the formation of compact skyrmions, affecting their stability and dynamics\cite{MantelPRB20}. Additionally, DMIs significantly modify spin-wave excitations in low-dimensional and confined geometries, leading to unique spin-wave modes and quantum entanglement effects in nano-rings and spin clusters\cite{nano14030286, Mahdavifar2024}.

\begin{figure}
    \centering
    \includegraphics[width=6cm]{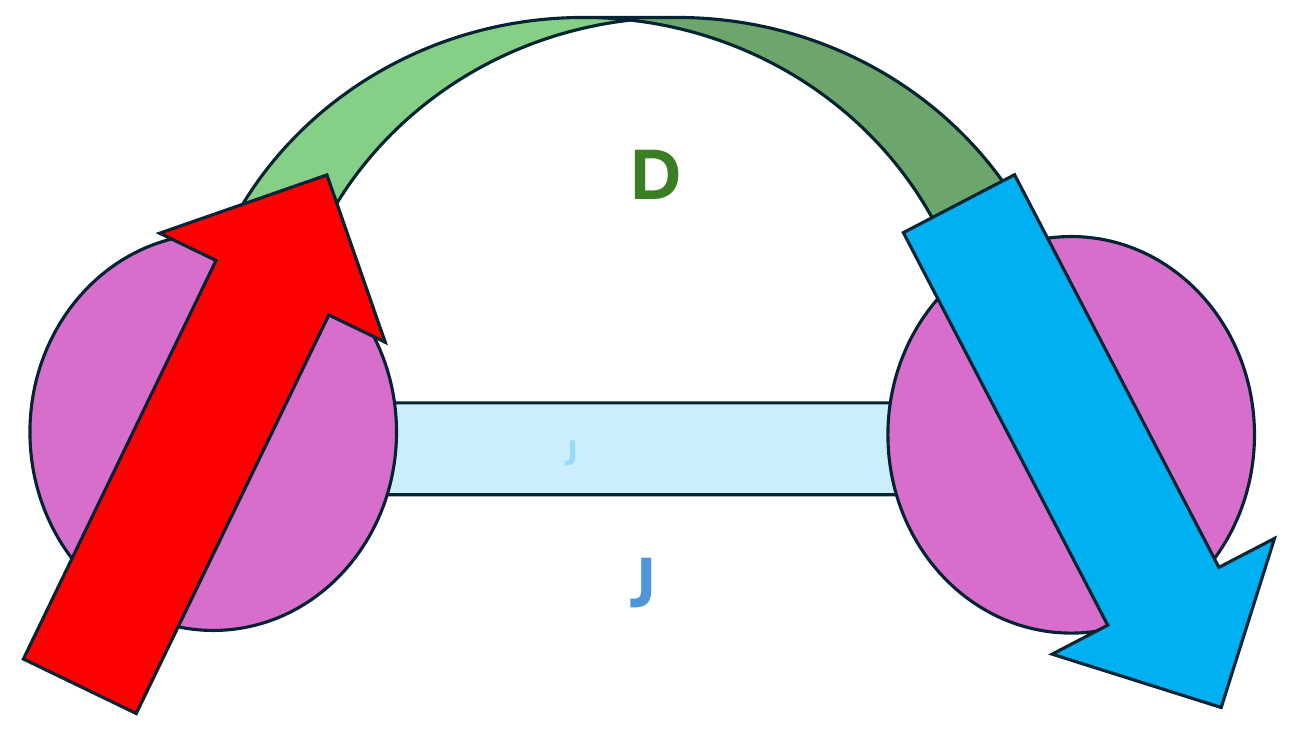}
    \caption{Illustration of the $S=\frac{1}{2}$ Heisenberg–Dzyaloshinskii–Moriya (HDM) quantum spin dimer. The general wavefunction $\ket{\psi}$ is a superposition of all spin configurations with real and complex amplitudes $a$, $b$, $c$, and $d$. The Heisenberg exchange interaction $J$ couples the spins isotropically (shown along the bond), while the antisymmetric Dzyaloshinskii–Moriya interaction $\mathbf{D}$ (green arc) introduces spin canting via a vector product term, typically arising from broken inversion symmetry between the sites.}
    \label{HDMdimerConfig}
\end{figure}

In ultrathin films, the presence of anisotropic interfacial DMIs has been shown to support the formation of not only skyrmions\cite{fert2013skyrmions} but also bimerons, extending the range of possible chiral magnetic textures\cite{UdalovPRB21}. First-principles studies have provided a deeper insight into the fundamental mechanisms of DMIs, revealing their microscopic origin at metal interfaces, such as Co/Pt, where spin-orbit coupling and structural inversion asymmetry drive the emergence of chiral interactions\cite{YangPRL15}. In one-dimensional quantum systems, the Heisenberg chain with DMIs exhibits critical behavior and factorization phenomena, which influence spin correlation functions and entanglement properties\cite{YIPRB19}. Beyond low-dimensional systems, DMIs also play a role in twisted magnets such as Cu$_2$OSe$O_3$, where a combination of exchange interactions and local spin canting leads to complex chiral structures\cite{CHIZHIKOV2015142}. The influence of DMIs on skyrmion physics has been extensively studied in both theoretical and applied contexts, with skyrmions emerging as stable, mobile spin configurations that hold promise for future spintronic devices\cite{Finocchio_2016}.


{A DMI is not restricted to bulk magnetic systems with long-range magnetic ordering. Spin clusters have also exhibited this rotational anisotropy in materials like Mo$_3$S$_7$(dmit)$_3$ \cite{Powell17PRB}, Na$_{12}$[Cu$_3$(SbW$_9$O$_{33}$)$_2$(H$_2$O)$_3$]$\cdot$46H$_2$O \cite{Iida11PRB}, and Cu$_2$(C$_5$H$_{12}$N$_2$)$_2$Cl$_4$ \cite{Miyahara07PRB}, where spin are effective isolated producing dimers and trimers. Magnetic clusters have garnered increased attention due to their potential applications in spin qubit systems and other technologies\cite{Meier03PRL}. Vargova et al. examined the theoretical underpinning of a 1/2-1 dimer and the potential for magnetocaloric effects\cite{Vargova2022}. However, there is a distinct need for more detailed experimental and theoretical work.}

Theoretical models of spin clusters often leverage network-based approaches, where interactions between spins can be mapped onto complex quantum properties, revealing novel entanglement patterns and dynamical phase transitions\cite{ChepuriCP2023,van2024mapping}. The modeling of neutron scattering has played a crucial role in probing the magnetic properties of molecular spin clusters, enabling direct measurements of exchange interactions, anisotropies, and quantum fluctuations in these systems\cite{Haraldsen05PRB,CHABOUSSANT2004E51}. Due to their finite size, these systems exhibit distinctive quantum effects that can be further investigated using spectroscopy or thermodynamic measurements\cite{NICOLAZZI20181060, Chilton2022ARMR}. By understanding the dynamics of magnetic interactions between ions, spin coherence can be manipulated for future quantum technologies\cite{BaylissSci20,niknam2022quantum}.

{In this study, we investigate the effect of DMI on various spin-1/2 clusters, with a particular focus on the spin dimer. We examine the general structure factor equation for an S = 1/2 dimer and show that the anisotropic ratio $D_z/|J|$ and complex phase $\phi$ of the DMI control the gap energy and intensity of the $|0,0\rangle \to |1,0\rangle$ transition. Our approach employs a simple yet effective Hamiltonian that incorporates both isotropic and anisotropic interactions, as well as the effects of an external magnetic field and an electric field. Our results demonstrate that the DM interaction shifts energy eigenstates by producing a level repulsion and significantly modifies the structure factor.}

\section{Heisenberg-Dzyaloshinskii-Moriya Model}\label{FullModel}

The Hamiltonian for the Heisenberg-Dzyaloshinskii-Moriya (H-DM) model is given by

\begin{equation}\label{Hamiltonian}
    H = J \sum_{\langle i,j \rangle} \vec{S}_i \cdot \vec{S}_j 
    + \sum_{\langle i,j \rangle} \vec{D}_{ij} \cdot (\vec{S}_i \times \vec{S}_j)  
    +E_{B} \sum_{i \neq j} S_{i}^{z},
\end{equation}

\noindent where $J$ is the Heisenberg exchange interaction, {with $J > 0$ favoring antiferromagnetic coupling and $J < 0$ favoring ferromagnetic coupling. The operators $\vec{S}_i$ and $\vec{S}_j$ represent spin degrees of freedom at sites $i$ and $j$, while $\vec{D}_{ij}$ is the Dzyaloshinskii-Moriya (DM) interaction vector,} which introduces antisymmetric exchange and promotes non-collinear spin arrangements. The final term represents the Zeeman energy shift, $E_b = g\mu_B B_z$, which appears as a diagonal offset for the fully polarized states $|\uparrow\uparrow\rangle$ and $|\downarrow\downarrow\rangle$.




With the inclusion of the Dzyaloshinskii-Moriya interaction, the Heisenberg Hamiltonian of a simple dimer is modified into (\ref{HDMI}), preserving the diagonal elements while introducing new contributions in the off-diagonal components.

\begin{equation}\label{HDMI}
H= \left[\begin{array}{cccc}
\frac{J}{4} & 0 & 0 & 0 
\\
 0 & -\frac{J}{4} & \frac{J}{2}-\frac{i \mathit{Dz} \,{\mathrm e}^{i \phi}}{2} & 0 
\\
 0 & \frac{J}{2}+\frac{i \mathit{Dz} \,{\mathrm e}^{i \phi}}{2} & -\frac{J}{4} & 0 
\\
 0 & 0 & 0 & \frac{J}{4} 
\end{array}\right]
\end{equation}

The eigenvalues and eigenvectors are obtained by diagonalizing the Hamiltonian and are shown in Table \ref{tab:dimer_mixing}. The inclusion of off-diagonal terms due to the Dzyaloshinskii-Moriya (DM) interaction leads to nontrivial modifications in the mixed spin states \(\ket{\uparrow\downarrow}\) and \(\ket{\downarrow\uparrow}\), while the fully aligned states \(\ket{\uparrow\uparrow}\) and \(\ket{\downarrow\downarrow}\) remain unchanged. 

\begin{table}[h]
\caption{Spin states $|S_{\text{tot}}, S_z\rangle$ and their Clebsch-Gordan coefficients for a spin-$\frac{1}{2}$ dimer with Dzyaloshinskii--Moriya interaction along the $z$-direction and isotropic Heisenberg interaction. The $|0,0\rangle$ and $|1,0\rangle$ states are DM-mixed superpositions, while $|1,\pm1\rangle$ are unmixed product states. {Here, $\phi$ is a phase angle between the spins.}}
\centering
\begin{tabular}{|c|p{3.5 cm}|c|}
\hline
\centering $|S_{\text{tot}}, S_z\rangle$ & \centering\arraybackslash Spin States & \centering\arraybackslash Energy \\ \hline

$|0,0\rangle$ & \centering\arraybackslash $\displaystyle \frac{D_z\, e^{i\phi} i - J}{\sqrt{D_z^2 + J^2}}$$|\downarrow\uparrow\rangle + |\uparrow\downarrow\rangle$ & -$\frac{J}{4}$ - $\frac{\sqrt{J^2+D_z^2}}{2}$ \\ \hline

$|1,-1\rangle$ & \centering\arraybackslash $|\downarrow\downarrow\rangle$ & $\frac{J}{4}$\\ \hline

$|1,0\rangle$ & \centering\arraybackslash $\displaystyle \frac{D_z\, e^{i\phi} i - J}{\sqrt{D_z^2 + J^2}}$$|\downarrow\uparrow\rangle -|\uparrow\downarrow\rangle$ & -$\frac{J}{4}$ + $\frac{\sqrt{J^2+D_z^2}}{2}$\\ \hline

$|1,1\rangle$ & \centering\arraybackslash $|\uparrow\uparrow\rangle$ & $\frac{J}{4}$ \\ \hline

\end{tabular}
\label{tab:dimer_mixing}
\end{table}

The DM interaction modifies only the \(m_z = 0\) states, affecting both the singlet \((s_t = 0)\) and the corresponding triplet component \((s_t = 1, m_z = 0)\). This results in a complex phase contribution and a shift in their relative mixing, while the fully polarized states remain unchanged. These modifications highlight the anisotropic nature of the DM interaction, which breaks inversion symmetry and introduces a preferred chirality in the spin system\cite{Ham2021}.

\section{Thermodynamic Observables}\label{ThermoObservables}

In any quantum system, heat capacity is essential for understanding spin interactions, quantum excitations, and phase transitions governing a given material\cite{Affronte99PRB}. The heat capacity is easily obtained from the {partition function (Z)}, which is given by

\begin{equation}
    C = k_B \beta^2 \frac{d^2}{d\beta^2} \ln(Z),
\end{equation}

\noindent spanning over all eigenstates \( N \), where \( E_i \) denotes the corresponding eigenvalues, with \( \beta = \frac{1}{k_B T} \), where \( k_B \) is the Boltzmann constant and \( T \) is the temperature. Observations are scrutinized by checking that the respective calculations accurately yield the correct entropy for a general spin system, which can then be compared to the expected zero-temperature entropy of the system by

\begin{equation}
S = \int_0^\infty \frac{C}{\beta}d\beta = k_B \ln\left( \frac{N}{N_0} \right),
\end{equation}

\noindent where $N$ represents the total dimension of the Hilbert space, and $N_0$ denotes the degeneracy of the ground state. For small temperature variations, entropy peaks align with increases in heat capacity. Analyzing how the ground state evolves with changes in system parameters provides insight into the relationship between heat capacity peaks, eigenvalue degeneracies, and quantum phase transitions \cite{Haraldsen11PRL}.

\subsection{Magnetic Field–Induced Transitions in a DM-Active Spin Dimer}\label{MagneticTransitions}

\begin{figure}[ht]
    \centering
    \includegraphics[width=8.5cm]{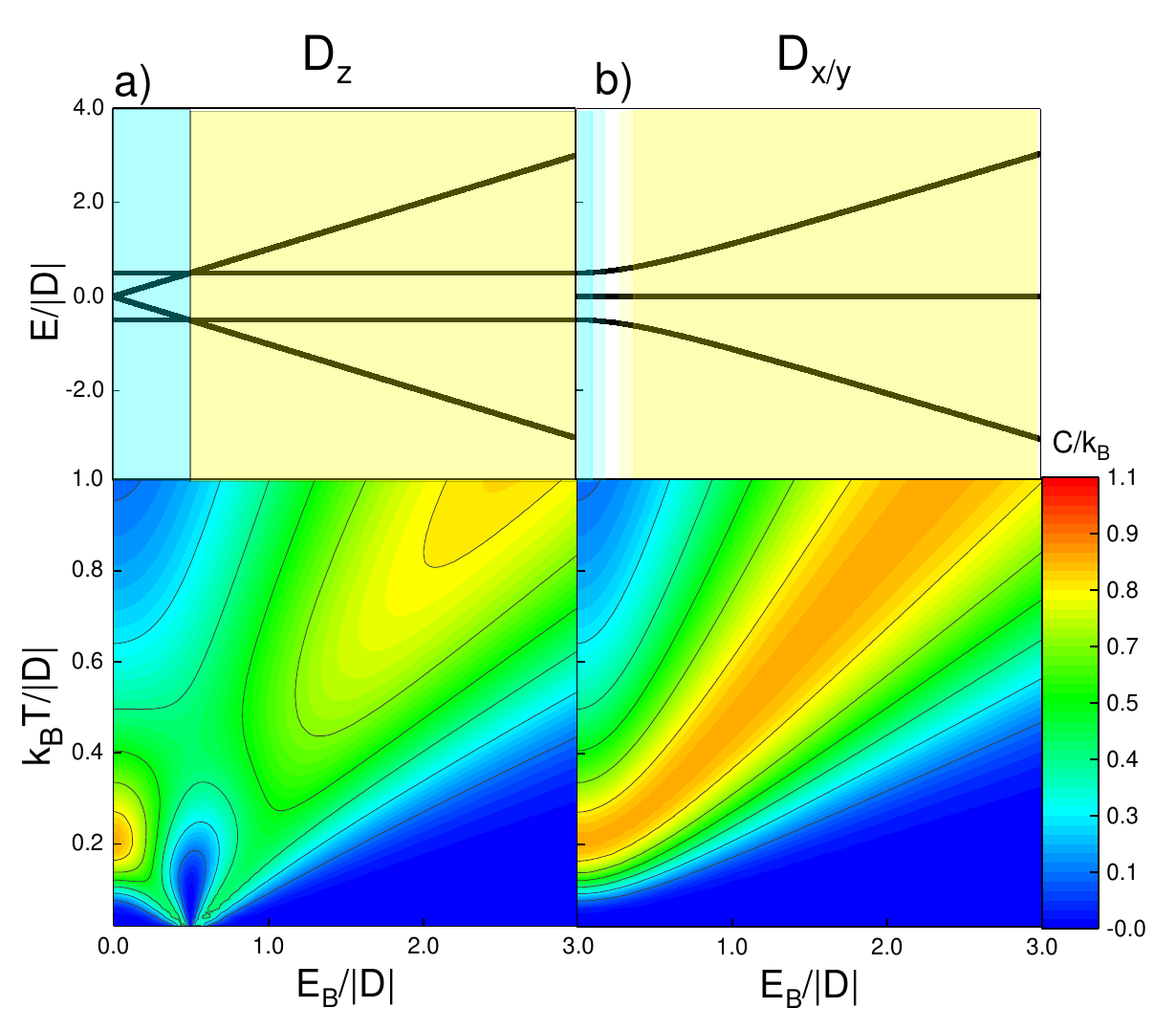}
    \caption{The energy levels from Table~\ref{tab:dimer_mixing}, plotted versus $E_B/|D|$ for a Dzyaloshinskii–Moriya $S = \tfrac{1}{2}$ dimer where $|J|$ = 0 and $|D|$ = 1, as the DM vector lies in the $z$ (a) plane and in the $xy$ plane (b). Corresponding contour plots show the reduced heat capacity $C/(k_B)$ versus $k_BT/|D|$ and $E_B/|D|$. For $\mathbf{D} \parallel z$, two $S_z = 0$ states remain field-independent, and the Schottky peak is set by a first-order phase transition from $|0,0\rangle (blue) \to |1,-1\rangle$ (yellow), which occurs at $E_B/|D| \approx 0.5$. In-plane $\mathbf{D}$ mixes all states and connects the two peaks featured in the $D_z$ case by a second-order phase transition from $|0,0\rangle \to |1,-1\rangle$. {The blue to yellow coloring in the energy level diagrams denotes a shift in the spin ground state.}
}
    \label{DMvsEb}
\end{figure}

Analyzing temperature and field dependance of the heat capacity provides distinct differences in how the magnetic interactions exchange information. Figure~\ref{DMvsEb} illustrates how a pure Dzyaloshinskii–Moriya (DM) interaction reshapes both the excitation spectrum and the reduced heat capacity of an S = 1/2 dimer. When the magnetic field is aligned with the DM vector, the Zeeman term commutes with the Hamiltonian, keeping the $S_z = 0$ states field-independent while the $\lvert1,\!\pm1\rangle$ levels shift linearly. A symmetry-protected crossing at $E_B/|D| \simeq 0.50$ causes the ground state to switch abruptly from the predominantly singlet $\lvert0,0\rangle$ to the fully polarized $\lvert1,-1\rangle$, releasing latent heat that manifests as a sharp vertical Schottky ridge in the heat-capacity map. Rotating the field into the $xy$ plane breaks $S_z$ conservation, allowing all four states to hybridize. The crossing is replaced by an avoided level repulsion, and the associated latent-heat feature broadens into a diagonal crest, signaling a continuous second-order quantum phase transition rather than a first-order jump.

\begin{figure}[ht]
    \centering
    \includegraphics[width=8.5cm]{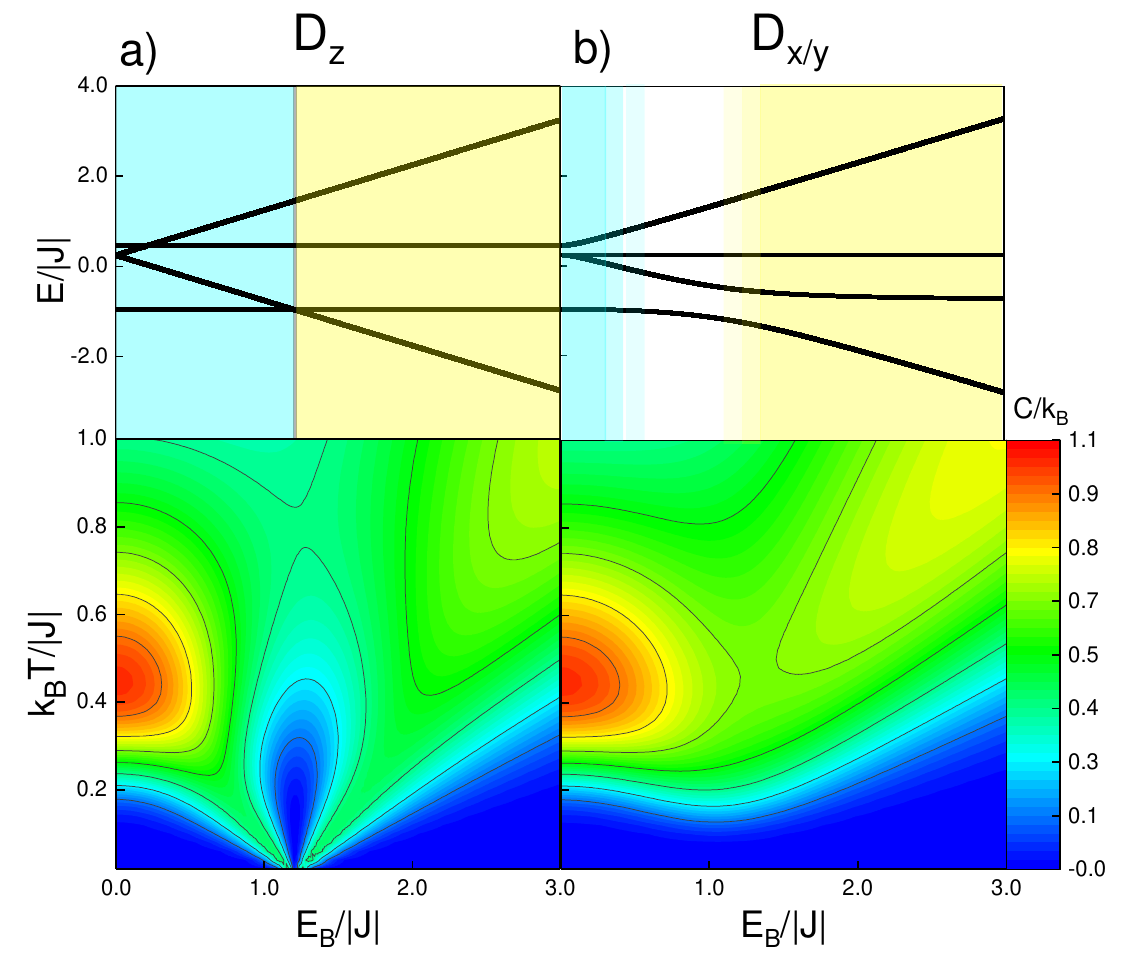}
    \caption{The energy levels from Table~\ref{tab:dimer_mixing}, plotted versus $E_B/|J|$ for a Heisenberg–DM $S = \tfrac{1}{2}$ dimer, where $|J|$ = $|D|$, as the DM vector is along $z$ (a) and in the $xy$ plane (b). Corresponding contour plots show the reduced heat capacity $C/(k_B)$ versus $k_BT/|J|$ and $E_B/|J|$. For $\mathbf{D} \parallel z$, the singlet–triplet gap is preserved, with the dominant Schottky peak set by the $|0,0\rangle$ (blue) $\to |1,-1\rangle$ (yellow) transition. In contrast, in-plane $\mathbf{D}$ mixes all triplet states and induces level repulsion, which produces a second-order phase transition from $|0,0\rangle \to |1,-1\rangle$, while broadening and shifting the heat-capacity anomaly. {The blue to yellow coloring in the energy level diagrams denotes a shift in the spin ground state.}}
    \label{HDMvsEb}
\end{figure}

Figure~\ref{HDMvsEb} extends this analysis to a dimer with added isotropic Heisenberg exchange $J$, taken here to be the same magnitude and sign as $D$.  The spectrum retains the flat $S_z = 0$ branches and linear $\lvert1,\!\pm1\rangle$ branches, but the crossing and the associated first-order transition shift to $E_B/|J| \simeq 1.2$, moving the Schottky anomaly accordingly. When the magnetic field is applied perpendicular to $\mathbf{D}$, the DM interaction again induces mixing within the triplet manifold, resulting in level repulsion and a broadened heat-capacity peak qualitatively similar to the case $D_z$, but now occurring at higher fields and without a distinct transition. The anomaly becomes an extended thermal ridge that shifts and smooths out as the temperature increases.

Together, these cases emphasize the complementary roles of $J$ and $D$, the Heisenberg exchange determines the base energy gap, while the orientation of the DM vector dictates whether the gap closes sharply or is softened by hybridization. The contrasting thermodynamic signatures of narrow versus broadened heat capacity anomalies offer a robust experimental tool for extracting both the magnitude and direction of the DM interaction in real quantum-spin dimers.

\subsection{Electric Field–Induced Quantum Criticality in a DM-Active Spin Dimer}\label{ElectricTransitions}

\begin{figure}[ht]
    \centering
    \includegraphics[width=8.5cm]{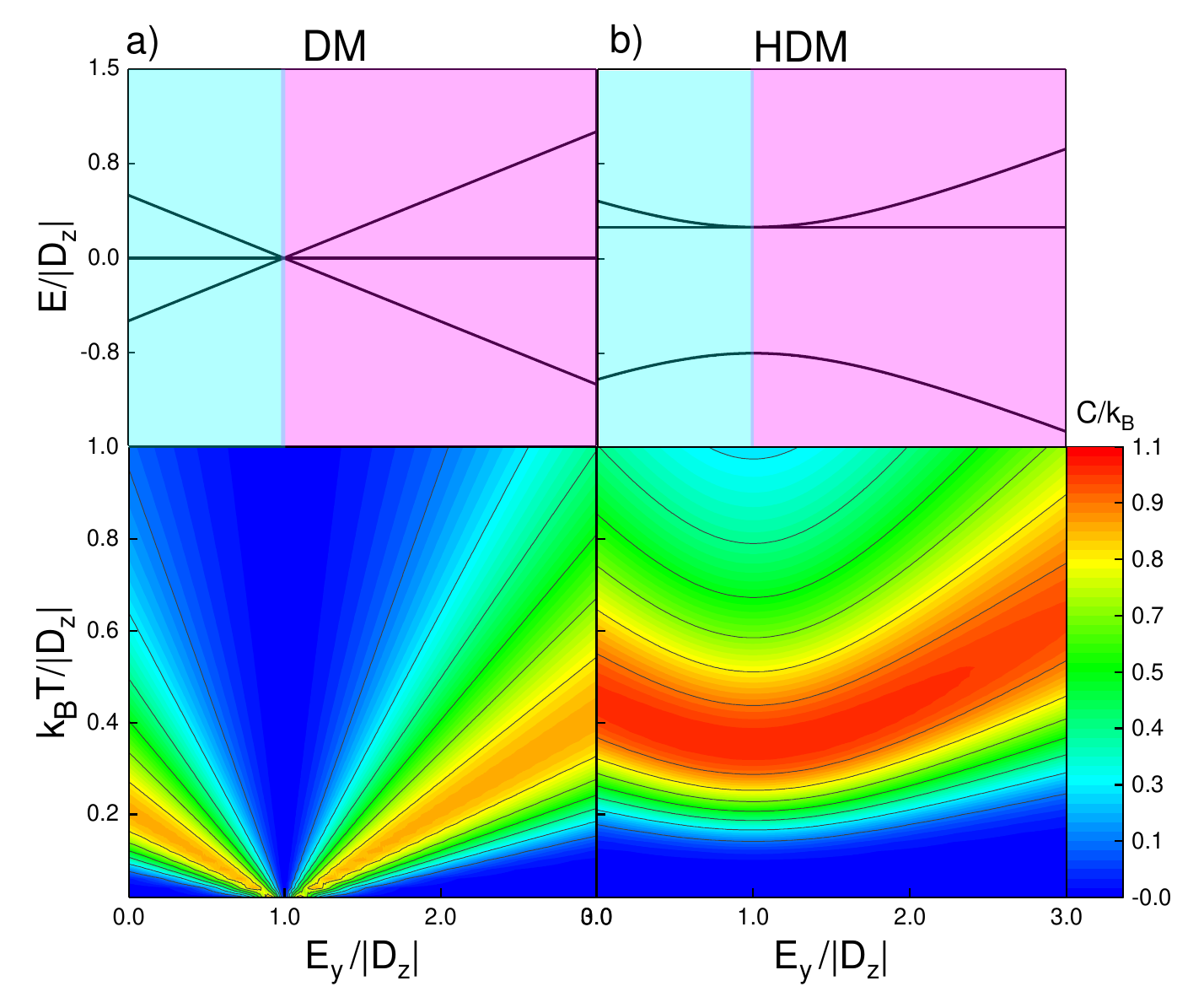}
    \caption{Energy levels and heat‑capacity contours for a spin‑$\tfrac12$ dimer as an in‑plane electric field $E_y$ tunes the DM component $D_z=D_{0}-\lambda E_y$. On the left $|J|$= 0 and $|D|$ = 1 (a), so pure DM levels cross linearly at $E_y/|D_z|=1$, giving a sharp fan‑like peak centered at a first order quantum phase transition from $|0,0\rangle$(blue) $\to |1,0\rangle$(violet). On the right (b), $|J|$ = $|D|$, the HDM crossing becomes avoided, and the heat‑capacity maximum broadens as the QPT becomes second order.}
    \label{HDMvsEy}
\end{figure}

Anisotropic interactions can induce spin canting, which gives rise to a local electric polarization via the Katsura– Nagaosa–Balatsky (KNB) mechanism. When an external electric field is applied, it couples to this polarization, resulting in an additional term in the Hamiltonian that modifies the spin interactions\cite{Vargova2022}. This field-induced contribution enables electric control of magnetic anisotropy and spin chirality.

We investigate a S = 1/2 dimer with Heisenberg and DM interactions, where the intrinsic DM vector is aligned along the $z$-axis, $\vec{D}_0 \parallel \hat{z}$, and the bond vector is $\vec{e}_{12} = \hat{x}$. Rather than an induced DMI by electric polarization, we apply an external electric field $\vec{E} \parallel \hat{y}$ that modifies an existing DM interaction via coupling to the spin-induced polarization described by the KNB mechanism. The system Hamiltonian is given by
\begin{equation}
    H = J\, \vec{S}_1 \cdot \vec{S}_2 + \vec{D}_\text{eff} \cdot (\vec{S}_1 \times \vec{S}_2) + E_B \cdot (S_{1}^z + S_{2}^z),
\end{equation}
where the effective DM vector depends on the electric field as
\begin{equation}
    \vec{D}_\text{eff} = \vec{D}_0 + \lambda (\vec{E} \times \vec{e}_{12}),
\end{equation}
which for $\vec{E} = E_y \hat{y}$ and $\vec{e}_{12} = \hat{x}$ simplifies to
\begin{equation}
    D_z^\text{eff} = D_0 - \lambda E_y.
\end{equation}

\noindent where $\lambda$ is the coupling strength of the DMI and the electric field.

This behavior is consistent with KNB theory, where the induced polarization
\begin{equation}
    \vec{P}_{ij} \propto \vec{e}_{ij} \times (\vec{S}_i \times \vec{S}_j)
\end{equation}
can be tuned by an external electric field. Depending on the symmetry of the system, the field may either modify an existing DM interaction or generate one where it is otherwise forbidden. This minimal spin dimer model illustrates how electric fields can control spin chirality and induce sharp ground-state crossovers in spin–orbit–coupled systems. The electric field can induce Schottky-type anomalies and level crossings that serve as precursors to quantum criticality in extended or frustrated magnetic systems.

\begin{table}[h]
\centering
\footnotesize               
\setlength{\tabcolsep}{3pt}    
\renewcommand{\arraystretch}{1.3}

\resizebox{\columnwidth}{!}{%
\begin{tabular}{|c|p{5.2cm}|c|}
\hline
\centering $|S_{\text{tot}},S_z\rangle$ &
\centering\arraybackslash Spin State &
\centering\arraybackslash Energy \\ \hline

$|0,0\rangle$ &
$\displaystyle
 \frac{\bigl((\lambda E_y-D_{0z})\,e^{i\phi}i-J\bigr)}
      {\sqrt{J^{2}+(\lambda E_y-D_{0z})^{2}}}\:
 |\!\downarrow\uparrow\rangle
 + |\!\uparrow\downarrow\rangle$ &
$-\dfrac{J}{4}-
 \dfrac{\sqrt{J^{2}+(\lambda E_y-D_{0z})^{2}}}{2}$ \\ \hline

$|1,-1\rangle$ & \centering
$|\!\downarrow\downarrow\rangle$ &
$\dfrac{J}{4}-E_B$ \\ \hline

$|1,0\rangle$ &
$\displaystyle
 \frac{\bigl((\lambda E_y-D_{0z})\,e^{i\phi}i-J\bigr)}
      {\sqrt{J^{2}+(\lambda E_y-D_{0z})^{2}}}\:
 |\!\downarrow\uparrow\rangle
 - |\!\uparrow\downarrow\rangle$ &
$-\dfrac{J}{4}+
 \dfrac{\sqrt{J^{2}+(\lambda E_y-D_{0z})^{2}}}{2}$ \\ \hline

$|1,1\rangle$ & \centering
$|\!\uparrow\uparrow\rangle$ &
$\dfrac{J}{4}+E_B$ \\ \hline
\end{tabular}}

\caption{Spin states and eigen‑energies of a spin‑$\tfrac12$ dimer whose bond vector is oriented along $\hat{\mathbf x}$.  
An intrinsic Dzyaloshinskii–Moriya vector $\mathbf D^{(0)} = D_{0z}\,\hat{\mathbf z}$ is linearly modulated by an electric field applied along $\hat{\mathbf y}$, yielding the field‑dependent antisymmetric exchange
$\mathbf D(E_y)=\mathbf D^{(0)} - \lambda E_y \,\hat{\mathbf z}$.  
A static longitudinal field $B\parallel\hat{\mathbf z}$ introduces the Zeeman shift $E_B = g\mu_B B$.}

\label{tab:dimer_mixing_Efield}
\end{table}

Examining the electric‑field dependence of the heat capacity uncovers a complementary route to disentangling exchange mechanisms.  Figure \ref{HDMvsEy} shows how an in‑plane electric field tunes the antisymmetric coupling via $D_z = D_{0}-\lambda E_{y}$ and thereby reshapes the excitation spectrum and thermal response of an $S=\tfrac12$ dimer.  In the pure DM limit, the singlet  $\lvert0,0\rangle$ and triplet $\lvert1,0\rangle$ branches vary linearly with $|D_z|$. When the field cancels the intrinsic DM term, at $E_{y}/|D_z|\simeq1$, these two levels cross, forcing the ground state to jump abruptly from $\lvert0,0\rangle$ to $\lvert1,0\rangle$.  The latent heat liberated at this first-order electric field transition appears as a sharp, needle‑like Schottky peak in the heat‑capacity map.

Introducing an isotropic Heisenberg exchange $J$ of the same magnitude and sign as $D$. This forces the linear crossing to be avoided and the minimum gap fixed by $J$. Consequently, the latent heat spike broadens into a smooth ridge that increases to higher $E_{y}/|J|$ and broadens with temperature, erasing any discrete transition. The resulting heat capacity anomaly resembles the broad crest produced by a transverse magnetic field in Fig.~\ref{HDMvsEb}, but here it is driven solely by electric field control of the DM interaction.

Taken together, these electric field results reinforce the complementary roles of $J$ and $D$.  The Heisenberg exchange defines the baseline singlet–triplet gap, whereas the tunable DM component determines whether that gap closes sharply or deforms continuously.  The contrast between a narrow Schottky spike and a broadened thermal ridge thus provides a practical experimental handle for extracting both the magnitude of $J$ and the electro-optic tunability of $D$ in real quantum spin dimers.

\subsection{DMI Signatures}\label{Signatures}

\begin{figure*}[ht]
    \centering
    \includegraphics[width=6in]{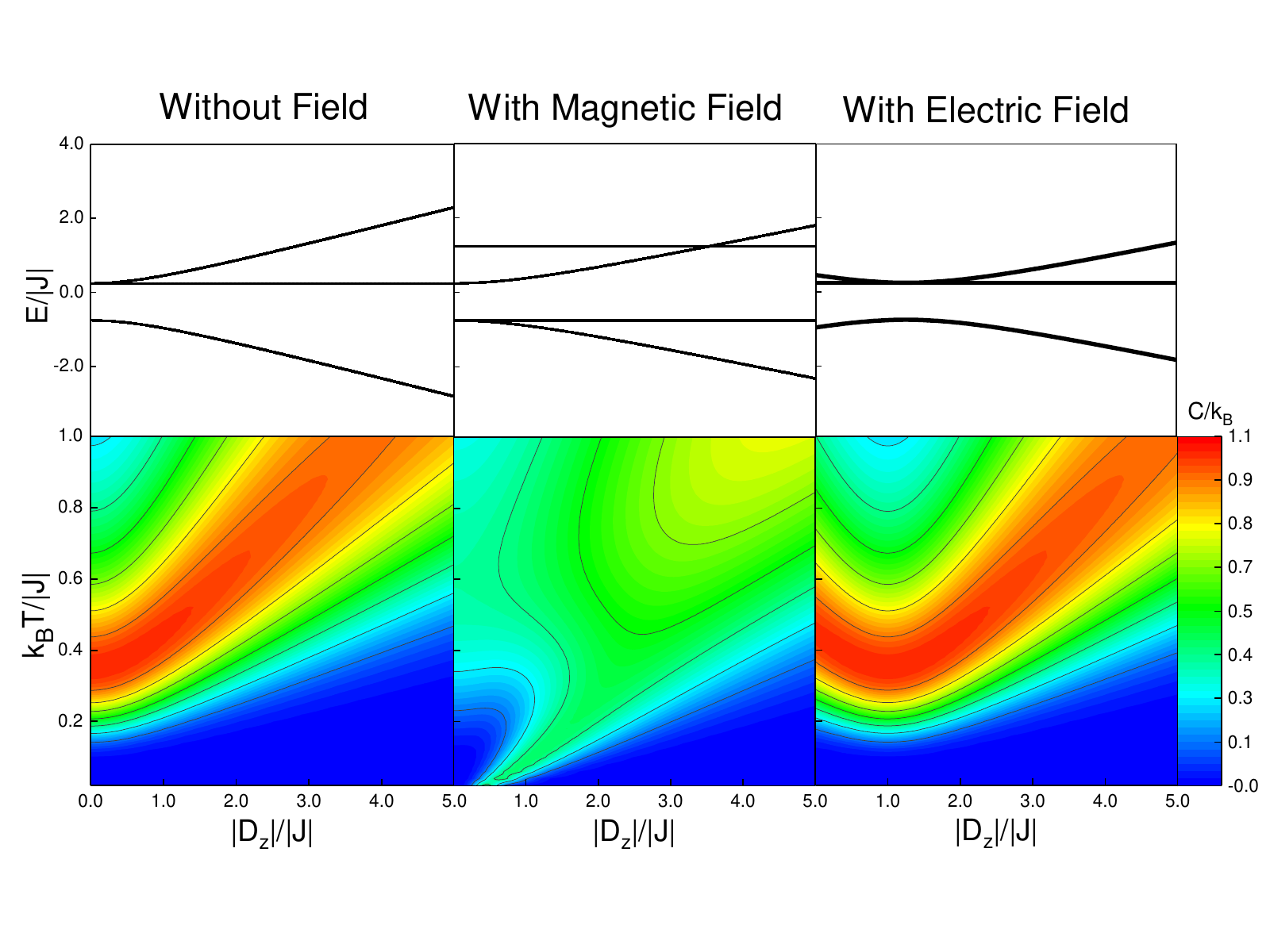}
    \caption{ Energy levels (top) and reduced heat capacity $C/(k_B)$ (bottom) of an $S = \tfrac{1}{2}$ Heisenberg–DM dimer as functions of $|D_z|/|J|$, shown for three cases: no external field (left), a longitudinal magnetic field (center), and a longitudinal electric field coupling to the DM vector (right). Without field, the singlet mixes with the $S_z = 0$ triplet, leading to a monotonic gap increase and a single Schottky ridge shifting to higher $T$. A magnetic field breaks the $\pm1$ triplet degeneracy and induces level repulsion between the $S_z = 0$ states, which splits and weakens the anomaly. The electric field distorts the spectrum without sharp anticrossings, maintaining a smaller singlet–triplet gap and restoring a broad, high-intensity heat-capacity peak.}
    \label{DMSignatures}
\end{figure*}

Scanning the out‑of‑plane DM strength provides a third vantage point on the balance between symmetric and antisymmetric exchange.  Figure~\ref{DMSignatures} follows how sweeping $|D_z|$ reshapes the spectrum and heat capacity of an S = 1/2 Heisenberg–DM dimer under three conditions, no external probe, a longitudinal magnetic field, and an in plane electric field that couples directly to the DM vector.  

In the field‑free case, only the singlet and $m\!=\!0$ triplet hybridize; their separation grows smoothly with $|D_z|$, producing a single Schottky ridge that drifts toward higher temperature as the gap widens. The non-linearity is due to the J-DM coupling. Adding $E_B\!\parallel\!z$ splits the $m\!=\!\pm1$ triplet pair. The descending $\lvert 1,-1\rangle$ branch approaches the lower $S_z\!=\!0$ level, level repulsion from the $DM-E_b$ coupling opens, and disrupts the DM driven Schottky anomaly, splitting the once‑coherent peak into weaker, fragmented features. This shows that Zeeman splitting redistributes entropy across several non-equivalent gaps. By contrast, an electric field leaves the $\pm1$ degeneracy intact but bends the $S_z\!=\!0$ manifold without a sharp anti-crossing, restoring a broad, high‑intensity thermal crest that sweeps diagonally across the $(|D_z|,T)$ plane.

These signatures show that each stimulus manipulates the same underlying singlet–triplet gap in a distinct way. The pure $|D_z|$ growth enlarges it monotonically, a longitudinal magnetic field fragments and quenches it, while an in-plane electric field reestablishes a strong Schottky anomaly by tuning the DM term itself.  The contrast among a drifting ridge, a split‑and‑weakened feature, and a restored broad peak therefore offers a clear experimental handle for disentangling Zeeman, electro‑optic, and intrinsic DM contributions in real quantum‑spin dimers.

\section{Neutron Scattering}

\subsection{Quantum Spin Dimer Structure Factor}

\begin{table*}[t]
    \centering
    \renewcommand{\arraystretch}{1.5}
    \begin{tabular}{|c|c|c|}
        \hline
        \textbf{Spin Operator} & \textbf{Overlap coefficients} & \textbf{Structure factor} \\
        \hline
        $\sigma_1^{+}e^{iqr_1}+\sigma_2^{+}e^{iqr_2}$ &
        $\begin{array}{l}
            \alpha_{+}=a_f^{\ast}c+b_f^{\ast}d \\[1pt]
            \beta_{+}=a_f^{\ast}b+c_f^{\ast}d
        \end{array}$ &
        $\displaystyle 
           S_{fi}^{+}(\mathbf q)=\eta_f^2\eta^2
           \Bigl[\,|\alpha_{+}|^{2}+|\beta_{+}|^{2}
            +2|\alpha_{+}||\beta_{+}|\cos(qr+\phi_{+})\Bigr]$ \\
        \hline
        $\sigma_1^{-}e^{iqr_1}+\sigma_2^{-}e^{iqr_2}$ &
        $\begin{array}{l}
            \alpha_{-}=c_f^{\ast}a+b_f^{\ast}d \\[1pt]
            \beta_{-}=b_f^{\ast}a+c_f^{\ast}d
        \end{array}$ &
        $\displaystyle 
           S_{fi}^{-}(\mathbf q)=\eta_f^2\eta^2
           \Bigl[\,|\alpha_{-}|^{2}+|\beta_{-}|^{2}
            +2|\alpha_{-}||\beta_{-}|\cos(qr+\phi_{-})\Bigr]$ \\
        \hline
        $\sigma_1^{z}e^{iqr_1}+\sigma_2^{z}e^{iqr_2}$ &
        $\begin{array}{l}
            \displaystyle
            \alpha_{z}=\frac12\!\bigl(a_f^{\ast}a+b_f^{\ast}b-c_f^{\ast}c-d_f^{\ast}d\bigr)\\[4pt]
            \displaystyle
            \beta_{z} =\frac12\!\bigl(a_f^{\ast}a-b_f^{\ast}b+c_f^{\ast}c-d_f^{\ast}d\bigr)
        \end{array}$ &
        $\displaystyle 
           S_{fi}^{z}(\mathbf q)=\eta_f^2\eta^2
           \Bigl[\,|\alpha_{z}|^{2}+|\beta_{z}|^{2}
            +2|\alpha_{z}||\beta_{z}|\cos(qr+\phi_{z})\Bigr]$ \\
        \hline
    \end{tabular}
    \caption{Overlap coefficients $\alpha_{\mu},\beta_{\mu}$ and corresponding
             structure‑factor expressions $S_{fi}^{\mu}(\mathbf q)$ for the
             transverse ($\mu=\pm$) and longitudinal ($\mu=z$) neutron‑scattering
             channels of a spin‑\,$\tfrac12$ dimer.  Initial‑state coefficients
             are $a,b,c,d$; $a_f,b_f,c_f,d_f$ belong to the final state.}
    \label{tab:GenStruc}
\end{table*}

Neutron scattering experiments probe the magnetic excitations of a system through the dynamical structure factor. For an S = 1/2 dimer, the scattering intensity captures the transitions between quantum states induced by momentum transfer \( \mathbf{q} \). Following the formalism developed by Haraldsen et al.~\cite{Haraldsen05PRB}, the general expression for the structure factor is given by

\begin{equation}
S_{ba}(\mathbf{q}) = \left| \langle \Psi_f | n_b^{o} | \Psi_i \rangle \right|^2,
\end{equation}

\noindent where \( n_{b}^{o} = \sigma^{\pm}_1 e^{i \mathbf{q} \cdot \mathbf{r}_1} + \sigma^{\pm}_2 e^{i \mathbf{q} \cdot \mathbf{r}_2} \) is the neutron magnetic scattering operator and \( \sigma_j^{\pm} = \sigma_j^x \pm i \sigma_j^y \) are the spin raising and lowering operators. The term \( e^{i \mathbf{q} \cdot \mathbf{r}_j} \) reflects the plane-wave nature of the neutron, with \( \mathbf{q} \) representing the momentum transfer and \( \mathbf{r}_j \) the position of the spin site, capturing the spatial modulation of the neutron’s interaction with the magnetic system. 

For a spin-$\frac{1}{2}$ dimer, we derive a general expression for the scattering intensity, given by

\begin{equation}
    S^{\mu}_{fi}(q)=\eta_f^2\eta_i^2[|\alpha_{\mu}|^{2}+|\beta_{\mu}|^{2}
                +2|\alpha_{\mu}||\beta_{\mu}|
                  \cos\!\bigl(q\vec{a} +\phi_{\mu}\bigr)],
\end{equation}

\noindent where $\alpha$ and $\beta$ represent the overlap between the initial and final eigenvector components provided in table \ref{tab:GenStruc}. The displacement \( \Delta r = r_2 - r_1=\vec{a} \) denotes the separation between the two sites in the dimer. The normalization factors \( \eta_f \) and \( \eta \) ensure proper normalization of the eigenvectors and are given by the inverse square root of the sum of the squared eigenvectors. The Clebsch–Gordan coefficients that appear in $\alpha$ and $\beta$ exclude the initial state $\ket{\uparrow\uparrow}$ and the final state $\ket{\downarrow\downarrow}$ when acting with the raising operator. By symmetry, the opposite is true for the lowering operator.

The site‑resolved overlaps $\alpha_{\mu}$ (site 1) and $\beta_{\mu}$ (site 2) measure how strongly the eigenstate pair couples to spin fluctuations on each site, their magnitudes setting the incoherent background
$|\alpha_{\mu}|^{2}+|\beta_{\mu}|^{2}$. The relative phase  $\phi_{\mu}= \phi_{\beta_{\mu}}-\phi_{\mu}$, quantifies the interference between the two sites. A finite Dzyaloshinskii–Moriya interaction renders the eigenvectors
complex and conjugates the raising and lowering overlaps, producing phases
of opposite sign, $\phi_{-}=-\phi_{+}.$ The resulting shift of the constructive‑interference maxima in $q$
provides a direct, phase‑sensitive probe of the chiral interaction’s sign and magnitude. Since \(\sigma_{1}^{+}\) flips spin‑down \(\!\to\!\) up on site 1 while
\(\sigma_{1}^{-}\) flips up \(\!\to\!\) down, the overlaps that enter
\(\alpha_{+}\) and \(\alpha_{-}\) are complex conjugates of each other
(and likewise for \(\beta_{\pm}\)).

As spins are added to the system with nearest-neighbor interactions, the site-resolved overlap coefficients grow combinatorially, capturing the contributions from all basis state transitions involving a single spin flip. The general form for the overlap coefficient on site \(j\) from the raising channel is

\begin{equation}
A_j
   = \sum_{u:\,s_j=0}
     {a^{(f)}_{\,u\oplus\hat\jmath}}^{\!*}\,a_{u},
\label{eq:Cj_def}
\end{equation}

where \(u = (s_1, s_2, \dots, s_n)\) is a binary string representing the spin configuration of the system in the computational basis, with \(s_j = 0\) indicating a down spin at site \(j\), and \(u \oplus \hat\jmath\) denotes the bitstring obtained by flipping the \(j^{\mathrm{th}}\) bit of \(u\) from 0 to 1. This expression applies generally to any \(n\)-site spin chain. We expand the general structure for N-spin systems, and establish overlap coefficients and structure factor equations for the trimer, tetramer, and pentamer in table \ref{tab:ExplicitOverlaps} and table  \ref{tab:StructureFactorExpansions} in the appendix.

\begin{figure}[ht]
    \centering
    \includegraphics[width=8cm]{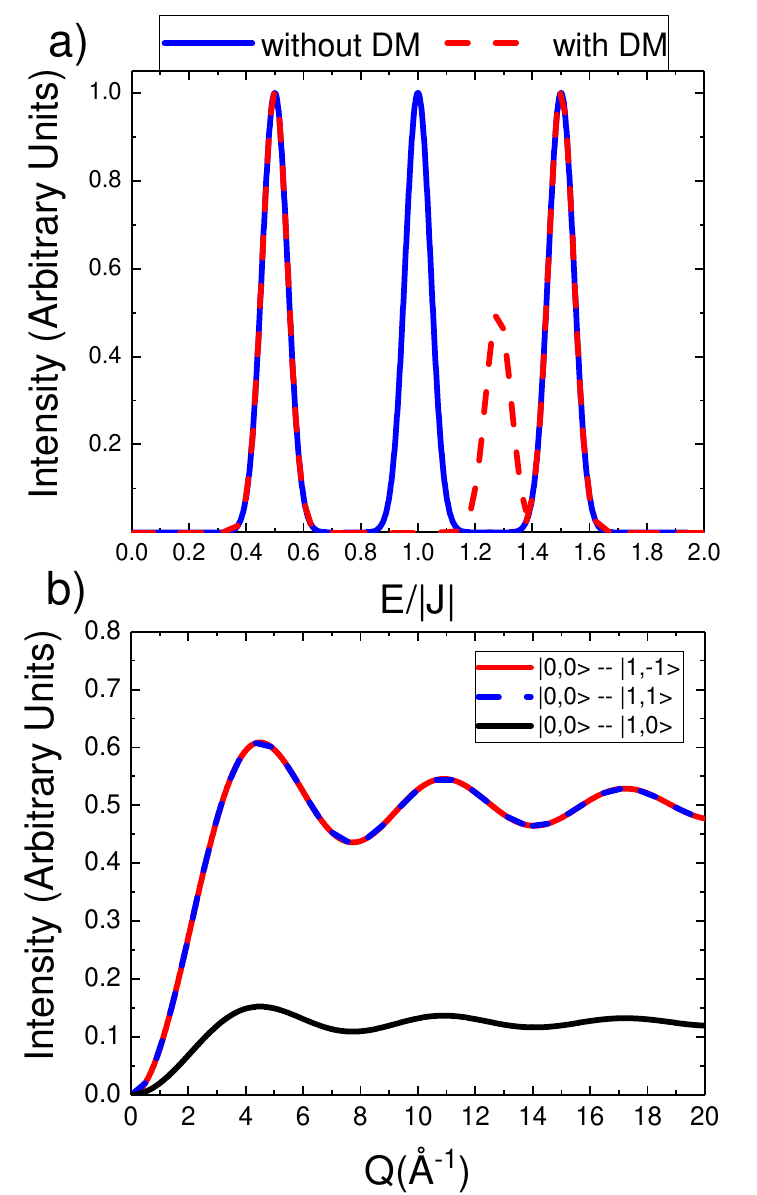}
    \caption{ Simulated neutron scattering intensity for a spin-$\tfrac{1}{2}$ Heisenberg–Dzyaloshinskii–Moriya (HDM) dimer under an external magnetic field, with $E_B = 0.5$ and $|J| = |D|$. (a) Intensity versus energy transfer $E/|J|$, where the leftmost peak corresponds to the transition $\Delta E_{-1} = |1,-1\rangle - |0,0\rangle$, the central peaks to $\Delta E_0 = |1,0\rangle - |0,0\rangle$, and the rightmost peak to $\Delta E_1 = |1,1\rangle - |0,0\rangle$. The DM interaction shifts the energies and its phase ($\phi_z$) redistributes the intensity, particularly affecting $\Delta E_0$. (b) Powder-averaged structure factor intensity as a function of momentum transfer $Q$. Spin-flip transitions $|0,0\rangle \to |1,\pm1\rangle$ remain largely unchanged, while the non-spin-flip transition $|0,0\rangle \to |1,0\rangle$ shows a clear reduction in amplitude and altered $Q$-dependence due to the DM interaction.}
    \label{IvsE}
\end{figure}

As the number of spins \(n\) in the cluster increases, the structure factor \(S_n(q)\) naturally decomposes into self-interference and pair-interference terms, reflecting the growing spatial correlations across the system. For a uniform chain with lattice spacing \(a\) and site-resolved overlaps \(A_j\), the total intensity is given by
\begin{equation}
S_n(q) = \sum_{j=1}^{n} |A_j|^2 + 2 \sum_{k=1}^{n-1} \left( \sum_{j=1}^{n-k} |A_j A_{j+k}| \right) \cos(kqa + \phi_{jk}),
\end{equation}
where \(\phi_{jk} = \phi_{j+k}- \phi_j\) captures the relative phase between amplitudes. The self-term grows linearly with \(n\), while the coherent forward-scattering peak at \(q = 0\) scales quadratically as \(\left( \sum_j A_j \right)^2\), reflecting Bragg-like constructive interference when site amplitudes are in phase. As the cluster grows, additional cosine terms corresponding to new pairwise separations (\(k = 2, 3, \dots, n-1\)) contribute oscillatory structure, and the main lobe of the intensity profile narrows proportionally to \(1/n\). In the special case of uniform real amplitudes, this expression reduces to a squared sinc envelope, \(S_n(q) \propto \mathrm{sinc}^2(nqa/2)\), exhibiting zeros at integer multiples of \(2\pi / na\). The presence of phase shifts \(\phi_{jk}\), such as those induced by Dzyaloshinskii–Moriya interactions or site asymmetry, leads to constructive interference maxima that deviate from integer \(q\)-values, enabling direct experimental access to local chiral interactions. This formulation highlights the scalable coherence of extended spin clusters while preserving analytic tractability and interpretability through the site-resolved amplitudes \(|A_j|\) and their relative phases.

{We illustrate the intensity spectrum of neutron scattering from a spin $\frac{1}{2}$ dimer in Figure \ref{IvsE}, where the energy spectra are calculated, along with the powder-averaged structure factor. We use the powder-average method, as it is a standard approach employed for inelastic neutron scattering, since many materials lack sufficient crystalline material for neutron scattering measurements.} Notably, the DM interaction shifts spectral weight and redistributes intensity, unveiling additional features that emerge due to the anisotropic exchange. The middle peak is reduced in intensity and moved to higher energies, while the other two peaks remain unaffected. The middle peak, corresponding to the \( |1,0\rangle - |0,0\rangle \) transition, is the only one affected because the Dzyaloshinskii-Moriya (DM) interaction along \( z \) (\( D_z \)) directly mixes the \( |1,0\rangle \) and \( |0,0\rangle \) states. This hybridization shifts their energy levels, altering the transition energy and leading to the observed peak shift. In contrast, the \( |1,1\rangle \) and \( |1,-1\rangle \) states are pure eigenstates of \( S^z \) and remain unaffected by \( D_z \), since the DM term only influences spin components perpendicular to \( z \). As a result, the corresponding transitions \( |1,1\rangle - |0,0\rangle \) and \( |1,-1\rangle - |0,0\rangle \) remain unchanged, as is supported by experimental work on quantum dimer systems\cite{SAKAI00JPSJ}. The associated unpolarized neutron structure factor is presented in Figure \ref{IvsE}.

The energy shift in the $\lvert S \rangle \leftrightarrow \lvert S_0 \rangle$ transition and a suppression of its structure factor due to the unequal overlap coefficients in the mixed eigenstate are shown in Table~\ref{tab:dimer_transitions}. Meanwhile, the $\lvert S \rangle \leftrightarrow \lvert S_{\pm1} \rangle$ transitions retain the same interference prefactor as in the pure Heisenberg case. This results in a distinctive neutron-scattering signature where a shifted and intensity-reduced central peak is  associated with $\lvert S_0 \rangle$, alongside nearly stationary and unchanged outer peaks for $\lvert S_{\pm1} \rangle$. Only for sufficiently large $D_z /|J|$ or improved resolution does the small quadratic shift of the outer peaks become resolvable.

\begin{table}[h]
\centering
\resizebox{\columnwidth}{!}{%
\begin{tabular}{|c|c|c|}
\hline
\textbf{Transition} & \textbf{Energy gap $\Delta E$} & \textbf{Structure factor $S(q)$} \\
\hline
$\lvert 0,0\rangle \rightarrow \lvert 1,-1\rangle$ &
$\displaystyle \frac{J}{2} + \frac{1}{2}\sqrt{J^{2}+D_{z}^{2}} + E_{B}$ &
$\displaystyle \frac{1}{2}[1+\cos(\vec{a}\cdot\vec{q})]$ \\
\hline
$\lvert 0,0\rangle \rightarrow \lvert 1,0\rangle$ &
$\displaystyle \sqrt{J^{2}+D_{z}^{2}}$ &
$\displaystyle \frac{|J|}{2\sqrt{J^{2}+D_{z}^{2}}}[1+\cos(aq+\phi)]$ \\
\hline
$\lvert 0,0\rangle \rightarrow \lvert 1,+1\rangle$ &
$\displaystyle \frac{J}{2} + \frac{1}{2}\sqrt{J^{2}+D_{z}^{2}} - E_{B}$ &
$\displaystyle \frac{1}{2}[1+\cos(\vec{a}\cdot\vec{q})]$ \\
\hline
\end{tabular}%
}
\caption{Energy gaps and structure-factor prefactors for neutron-induced transitions from the ground state $\lvert 0,0\rangle$ of a spin-$\tfrac{1}{2}$ dimer with a $D_{z}$ Dzyaloshinskii–Moriya interaction.}
\label{tab:dimer_transitions}
\end{table}

\begin{table*}[t]
\centering
\renewcommand{\arraystretch}{1.33}
\begin{tabular}{|c|>{\centering\arraybackslash}p{0.79\linewidth}|}
\hline
\textbf{System} & \textbf{Explicit overlap coefficients (raising channel, $n^{+}$)} \\ \hline\hline
Trimer ($n=3$) &
\[
\begin{aligned}
\alpha &= a_f^{*}e + f_f^{*}d + b_f^{*}g + c_f^{*}h,\\[4pt]
\beta  &= b_f^{*}c + e_f^{*}d + h_f^{*}f + g_f^{*}h,\\[4pt]
\Gamma &= a_f^{*}b + e_f^{*}g + e_f^{*}h + f_f^{*}c.
\end{aligned}
\] \\ \hline
Tetramer ($n=4$) &
\[
\begin{aligned}
\alpha &= a_f^{*}e + b_f^{*}f + b_f^{*}k + d_f^{*}n
       + j_f^{*}l + l_f^{*}p + m_f^{*}h + o_f^{*}g,\\[4pt]
\beta  &= a_f^{*}d + b_f^{*}l + c_f^{*}j + e_f^{*}n
       + f_f^{*}p + g_f^{*}h + k_f^{*}p,\\[4pt]
\Gamma &= a_f^{*}c + b_f^{*}o + d_f^{*}j + f_f^{*}g
       + k_f^{*}g + n_f^{*}l + p_f^{*}h,\\[4pt]
\zeta  &= a_f^{*}b + c_f^{*}o + d_f^{*}l + e_f^{*}f
       + e_f^{*}k + l_f^{*}h + j_f^{*}m + n_f^{*}p.
\end{aligned}
\] \\ \hline
Pentamer ($n=5$) &
\[
\begin{aligned}
\alpha &= a_f^{*}b + c_f^{*}d + e_f^{*}f + g_f^{*}h
       + i_f^{*}j + k_f^{*}l + m_f^{*}n + o_f^{*}p \\[-2pt]
       &\quad + q_f^{*}r + s_f^{*}t + u_f^{*}v + w_f^{*}x
       + y_f^{*}z + \mathrm{aa}_f^{*}\mathrm{ab}
       + \mathrm{ac}_f^{*}\mathrm{ad} + \mathrm{ae}_f^{*}\mathrm{af},\\[6pt]
\beta  &= a_f^{*}c + b_f^{*}d + e_f^{*}g + f_f^{*}h
       + i_f^{*}k + j_f^{*}l + m_f^{*}o + n_f^{*}p \\[-2pt]
       &\quad + q_f^{*}s + r_f^{*}t + u_f^{*}w + v_f^{*}x
       + y_f^{*}\mathrm{aa} + z_f^{*}\mathrm{ab}
       + \mathrm{ac}_f^{*}\mathrm{ae} + \mathrm{ad}_f^{*}\mathrm{af},\\[6pt]
\Gamma &= a_f^{*}e + b_f^{*}f + c_f^{*}g + d_f^{*}h
       + i_f^{*}m + j_f^{*}n + k_f^{*}o + l_f^{*}p \\[-2pt]
       &\quad + q_f^{*}u + r_f^{*}v + s_f^{*}w + t_f^{*}x
       + y_f^{*}\mathrm{ac} + z_f^{*}\mathrm{ad}
       + \mathrm{aa}_f^{*}\mathrm{ae} + \mathrm{ab}_f^{*}\mathrm{af},\\[6pt]
\zeta  &= a_f^{*}i + b_f^{*}j + c_f^{*}k + d_f^{*}l
       + e_f^{*}m + f_f^{*}n + g_f^{*}o + h_f^{*}p \\[-2pt]
       &\quad + q_f^{*}y + r_f^{*}z + s_f^{*}\mathrm{aa} + t_f^{*}\mathrm{ab}
       + u_f^{*}\mathrm{ac} + v_f^{*}\mathrm{ad}
       + w_f^{*}\mathrm{ae} + x_f^{*}\mathrm{af},\\[6pt]
\eta   &= a_f^{*}q + b_f^{*}r + c_f^{*}s + d_f^{*}t
       + e_f^{*}u + f_f^{*}v + g_f^{*}w + h_f^{*}x \\[-2pt]
       &\quad + i_f^{*}y + j_f^{*}z + k_f^{*}\mathrm{aa} + l_f^{*}\mathrm{ab}
       + m_f^{*}\mathrm{ac} + n_f^{*}\mathrm{ad}
       + o_f^{*}\mathrm{ae} + p_f^{*}\mathrm{af}.
\end{aligned}
\]
\\ \hline
\end{tabular}
\caption{Explicit Clebsch–Gordan overlap combinations for the $n^{+}$ (raising) transition in linear spin clusters of three, four, and five sites.  Roman letters $a,b,c,\dots$ denote amplitudes of the initial eigenvector; a subscript $f$ marks amplitudes in the final eigenvector.  Double‑letter labels $\mathrm{aa}$–$\mathrm{af}$ continue the binary enumeration beyond $z$.}
\label{tab:ExplicitOverlaps}
\end{table*}
\subsection{Structure Factors of Finite Spin Clusters and Closed Chains}

\begin{table*}[t]
\centering
\renewcommand{\arraystretch}{1.25}
\begin{tabular}{|c|>{\centering\arraybackslash}p{0.26\linewidth}|>{\centering\arraybackslash}p{0.60\linewidth}|}
\hline
\textbf{System} &
\textbf{Overlap set (from Table~\ref{tab:ExplicitOverlaps})} &
\textbf{Structure factor $S_n(q)$} \\ \hline\hline
Trimer &
$\{\alpha,\beta,\Gamma\}$ &
\scriptsize$
\begin{aligned}
S_3(q)=
 &\,|\alpha|^{2}+|\beta|^{2}+|\Gamma|^{2}\\
 &+2|\alpha||\beta|\cos(qa+\phi_{\alpha\beta})\\
 &+2|\beta ||\Gamma|\cos(qa+\phi_{\beta\Gamma})\\
 &+2|\alpha||\Gamma|\cos(2qa+\phi_{\alpha\Gamma})
\end{aligned}$ \\ \hline
Tetramer &
$\{\alpha,\beta,\Gamma,\zeta\}$ &
\scriptsize$
\begin{aligned}
S_4(q)=
 &\,|\alpha|^{2}+|\beta|^{2}+|\Gamma|^{2}+|\zeta|^{2}\\
 &+2\!\Bigl[|\alpha||\beta|\cos(qa+\phi_{\alpha\beta})
            +|\beta||\Gamma|\cos(qa+\phi_{\beta\Gamma})
            +|\Gamma||\zeta|\cos(qa+\phi_{\Gamma\zeta})\Bigr]\\
 &+2\!\Bigl[|\alpha||\Gamma|\cos(2qa+\phi_{\alpha\Gamma})
            +|\beta ||\zeta |\cos(2qa+\phi_{\beta\zeta})\Bigr]\\
 &+2|\alpha||\zeta|\cos(3qa+\phi_{\alpha\zeta})
\end{aligned}$ \\ \hline
Pentamer &
$\{\alpha,\beta,\Gamma,\zeta,\eta\}$ &
\scriptsize$
\begin{aligned}
S_5(q)=
 &\,|\alpha|^{2}+|\beta|^{2}+|\Gamma|^{2}+|\zeta|^{2}+|\eta|^2+2\bigl[|\alpha||\beta|\cos(qa+\phi_{\alpha\beta})\\
 &
          +|\beta||\Gamma|\cos(qa+\phi_{\beta\Gamma})
          +|\Gamma||\zeta|\cos(qa+\phi_{\Gamma\zeta})
          +|\zeta||\eta|\cos(qa+\phi_{\zeta\eta})\bigr]\\
 &+2\bigl[|\alpha||\Gamma|\cos(2qa+\phi_{\alpha\Gamma})
          +|\beta ||\zeta |\cos(2qa+\phi_{\beta\zeta})
          +|\Gamma||\eta |\cos(2qa+\phi_{\Gamma\eta})\bigr]\\
 &+2\bigl[|\alpha||\zeta|\cos(3qa+\phi_{\alpha\zeta})
          +|\beta ||\eta  |\cos(3qa+\phi_{\beta\eta})\bigr]\\
 &+2|\alpha||\eta|\cos(4qa+\phi_{\alpha\eta})
\end{aligned}$ \\ \hline
\end{tabular}
\caption{Momentum‑dependent structure factors obtained by inserting the
explicit overlap coefficients from Table~\ref{tab:ExplicitOverlaps} into
the general formula
$S_n(q)=\sum_j|A_j|^{2}+2\!\sum_{k}\bigl(\sum_{j}|A_jA_{j+k}|\bigr)
\cos(kqa+\phi_{jk})$.  The phases
$\phi_{jk}=\phi_{j+k}-\phi_{j}$ account for DM‑induced complex
overlaps.}
\label{tab:StructureFactorExpansions}
\end{table*}

The progression from trimer to pentamer in Table \ref{tab:ExplicitOverlaps} and Table \ref{tab:StructureFactorExpansions} foreshadows the behavior of a much longer ring. Every new spin introduces another spin-flip amplitude, and through its products with the existing ones, it brings an additional harmonic into the momentum profile. In the trimer, the three coefficients $\alpha$, $\beta$, $\Gamma$ generate only the fundamental and first overtone, so the intensity will be broad and two-lobed. The tetramer adds $\zeta$; the mixed terms $\alpha\zeta$ and $\beta\zeta$ are enough to narrow the central peak. The pentamer contributes $\eta$ and with it the fourth harmonic, further squeezing the intensity and shifting the outer maxima to larger wave vectors. The line shape, therefore, evolves toward the envelope of a finite crystallite, and the relative heights of successive harmonics measure how far the excitation spreads across the cluster.

The same bookkeeping is carried over to an $n$-site ring. Each coefficient $A_{j}$ is still the overlap of the initial and final eigenstates linked by a single spin raise at site $j$, and the full structure factor is the sum of all self-terms plus the cosine-weighted interference between every pair of sites. Every microscopic interaction, whether Heisenberg, Kitaev, uniform, or staggered Dzyaloshinskii-Moriya, appears only through the magnitudes $\lvert A_{j}\rvert$ and the phase differences $\phi_{j+k}-\phi_{j}$.

If the chain is translationally invariant and the eigenstates carry a Bloch momentum $k_{0}$, all amplitudes share a common magnitude $\lvert A\rvert$ and a phase $\phi_{j}=k_{0}aj$. The inner sum in the expression of the structure factor then delivers $n$ identical contributions and collapses into the diffraction form,
\[
S^{\text{ring}}_{n}(q)=
n\,\lvert A\rvert^{2}\,
      \frac{\sin^{2}\!\bigl[\tfrac{n}{2}(q+k_{0})a\bigr]}
           {\sin^{2}\!\bigl[\tfrac{1}{2}(q+k_{0})a\bigr]},
\]
with Bragg maxima at $q=-k_{0}+2\pi m/a$. A uniform Dzyaloshinskii–Moriya vector whose pitch is $\phi$ simply shifts the effective Bloch momentum to $k_{0}+\phi/a$ and displaces every peak by the same amount\cite{Karimi11PRB}. A staggered DM pattern destroys the single-phase relation; therefore, one must evaluate the general expression with the explicit set $\{\lvert A_{j}\rvert,\phi_{j}\}$ obtained from the numerical eigenvectors\cite{Oshikawa97PRL,Oshikawa99PRB}. In all cases the self-terms fix the overall intensity scale, the pair products build the interference fringes, and the phase differences expose any chiral twist imposed by the underlying interactions, so the algebra that governs table table \ref{tab:ExplicitOverlaps} and \ref{tab:StructureFactorExpansions} scales smoothly from trimer and tetramer to an arbitrarily long chain.

\section{Conclusions}

By deriving a general structure factor for the $S=\frac{1}{2}$ dimer, we effectively demonstrate that thermodynamic anomalies observed in heat capacity align directly with the spin-resolved selection rules in Figure~\ref{IvsE}. We show that the DM induces singlet–triplet hybridization, which partially decouples spin–orbit coupling, revealing a regime similar to the incomplete Paschen–Back effect that enables transitions forbidden in a pure Heisenberg model \cite{SAKAI00JPSJ,Ohta23JPSJ}. The non‑linearity intensifies with the anisotropy ratio $D_z/|J|$ and the DMI complex phase $\phi$ suppresses the intensity of the central transition, while its magnitude $|D|$ shifts the associated gap to higher energy, without requiring additional anisotropic terms\cite{Plumb2016QuasiparticleContinuum}.

 
Each Schottky anomaly in the heat capacity corresponds to the spectral weight in the inelastic neutron spectrum. Without DMI, the system yields symmetric triplet levels and a broad single anomaly. The introduction of $D_z$ mixes the $|0,0\rangle$ and $|1,0\rangle$ states, sharpening the Schottky peak and shifting the associated neutron line upward in energy, while also reducing its intensity through the suppression of overlap produced by the phase factor and magnitude ($J / 2\sqrt{J^2 + D_z^2}$).

External fields modulate this hybridized gap in different ways. A longitudinal magnetic field lifts the $|1,\pm1\rangle$ degeneracy, causing a first-order quantum phase transition from $|0,0\rangle$ to $|1,\pm1\rangle$. The magnetic field induces level repulsion in the x-y plane and eliminates a distinct transition, forcing the transition to be a second-order transition. Meanwhile, an in-plane electric field alters $D_z$ smoothly, tuning the $|0,0\rangle \leftrightarrow |1,0\rangle$ gap with a sign flip occurring in the pure DM, and sharp anti-crossings in the HDM case both around a transition from $|0,0\rangle$ to $|1,0\rangle$.

We find that singlet-triplet hybridization via DMI governs both macroscopic and microscopic observables. A magnetic field activates spin-flip transitions $|0,0\rangle \rightarrow |1,\pm1\rangle$, while an electric field selectively couples to the non-spin-flip transition $|0,0\rangle \rightarrow |1,0\rangle$. Together, these parameters reproduce the full triplet structure in neutron scattering and map directly onto the heat capacity landscape, offering a coherent experimental signature of the DM interaction where the energy and intensity of only one transition are affected.

{While the DMI does account for the cross-product of all components. The exchange interaction in this model is kept isotropic. Further analysis would be beneficial for determining whether an XXZ model or a Kitaev-like model is more suitable for these systems. However, more systems have to be identified before these models can be refined. Therefore, the isotropic model is a good first-order approximation while further investigation is ongoing. }

\section{Appendix}

The full spin $\frac{1}{2}$ dimer Hamiltonian includes isotropic Heisenberg exchange, Dzyaloshinskii–Moriya (DM) interactions along all three spatial directions, and a Zeeman term due to an external magnetic field, shown in Eq.\ref{Full_HamiltonianMatrix}. In the standard basis $\{ |\uparrow\uparrow\rangle, |\uparrow\downarrow\rangle, |\downarrow\uparrow\rangle, |\downarrow\downarrow\rangle \}$, the Heisenberg interaction contributes symmetric off-diagonal exchange terms between opposite-spin states, while the DM interaction introduces complex-valued antisymmetric terms that couple spin-flip processes with a bond-direction-dependent phase $e^{i\phi}$. The matrix elements proportional to $D_x$, $D_y$, and $D_z$ reflect the projections of the DM vector $\mathbf{D} = (D_x, D_y, D_z)$, with real and complex contributions from the $x$ and $y$ components and complex terms arising from $D_z$.  This generalized Hamiltonian accounts for spin-anisotropic exchange, enabling a direct investigation of the role of DM interactions and their symmetry-breaking effects on the energy spectrum and scattering amplitudes.

The DM vector $\mathbf D = D_{z}\hat{\mathbf z}$ is perpendicular to the bond, so the antisymmetric term couples only the singlet 
$\lvert S\rangle$ and the $m_{z}=0$ triplet $\lvert S_{0}\rangle$.  
Diagonalising this two‑state subspace now yields
\begin{equation}
E_{\pm} = -\frac{J}{4} \pm \frac{1}{2} \sqrt{J^{2}+D_{z}^{2}},
\end{equation}
while the $m_{z} = \pm 1$ triplet levels remain exact eigenstates at
\begin{equation}
E_{\pm 1} = +\frac{J}{4} \mp E_B,
\end{equation}
because the DM operator has no matrix element in that sector.  
Consequently, the singlet–$S_{0}$ gap is
\begin{equation}
\Delta E_0 = \sqrt{J^{2}+D_{z}^{2}},
\end{equation}
whereas the $\lvert S_{\pm 1} \rangle \rightarrow \lvert S \rangle$ gaps change only at second order, under a Taylor expansion we see, 
\begin{equation}
\Delta E_{\pm 1} = J + \frac{D_{z}^{2}}{2J} + \mathcal{O}(D_{z}^{4}).
\end{equation}

The DM mixing alters the interference prefactor in the $q$‑dependent part of the structure factor. Writing the ground‑state wavefunction as 
$\lvert \Psi_{0} \rangle = \alpha \lvert \downarrow \uparrow \rangle + \beta \lvert \uparrow \downarrow \rangle$, 
one finds
\begin{equation}
|\alpha||\beta| = \frac{|J|}{2\sqrt{J^{2}+D_{z}^{2}}} < \frac{1}{2} \quad (D_z \neq 0),
\end{equation}
so the amplitude of the singlet–$S_{0}$ peak is suppressed relative to the pure‑Heisenberg value, whereas the outer $m=\pm1$ peaks retain the Heisenberg prefactor $1/2$.  
In simulated (or experimental) inelastic‑neutron spectra, this combination of a central‑peak energy shift and intensity reduction, alongside essentially stationary outer peaks for moderate $D_{z}/J$, provides a clear signature of Dzyaloshinskii–Moriya interactions. Only for sufficiently large $|D_{z}|$ or finer energy resolution does the small quadratic shift of the $m=\pm1$ channels become observable.

\begin{equation}\label{Full_HamiltonianMatrix}
H = \left[
\begin{array}{cccc}
\frac{J}{4} + E_b & \frac{i D_x}{4} + \frac{D_y}{4} & -\frac{i D_x}{4} - \frac{D_y}{4} & 0 \\[8pt]
-\frac{i D_x}{4} + \frac{D_y}{4} & -\frac{J}{4} & \frac{J}{2} + \frac{iD_z \, e^{i \phi}}{2} & \frac{i D_x}{4} + \frac{D_y}{4} \\[8pt]
\frac{i D_x}{4} - \frac{D_y}{4} & \frac{J}{2} - \frac{i D_z \, e^{i \phi}}{2} & -\frac{J}{4} & -\frac{iD_x}{4} - \frac{D_y}{4} \\[8pt]
0 & -\frac{iD_x}{4} + \frac{D_y}{4} & \frac{i D_x}{4} - \frac{D_y}{4} & \frac{J}{4} - E_b
\end{array}
\right]
\end{equation}

\printcredits

\bibliographystyle{unsrtnat}
\bibliography{DMI-General-Reviewed1}

\end{document}